\documentclass[a4paper,11pt]{article}
\usepackage{jheppub}

\usepackage[utf8]{inputenc}
\usepackage{graphicx,cancel,float}
\usepackage{subfigure}
\usepackage{amssymb}
\usepackage{textcomp}
\usepackage{amsmath}
\usepackage{tabularx}
\usepackage{bm}
\usepackage{times}
\usepackage{color}
\usepackage{slashed}
\usepackage{multirow}
\usepackage{verbatim}
\usepackage{epsfig,epstopdf}

\allowdisplaybreaks[4]

\def\lsim{\mathrel{\raise.3ex\hbox{$<$\kern-.75em\lower1ex\hbox{$\sim$}}}}
\def\gsim{\mathrel{\raise.3ex\hbox{$>$\kern-.75em\lower1ex\hbox{$\sim$}}}}

\definecolor{orange}{rgb}{1,0.5,0}

\newcommand{\minigraph}[5][0.25in]{\begin{minipage}{#2}\begin{center}\includegraphics[width=#2]{#5}\\\vspace{#3}\hspace{#1}{\footnotesize #4}\end{center}\end{minipage}}

\preprint{}
\subheader{\it \today}

\title{Loop effect in the coherent neutrino-nucleus scattering}

\author[a]{Tong Li}
\emailAdd{litong@nankai.edu.cn}
\affiliation[a]{School of Physics, Nankai University, Tianjin 300071, China}
\author[b]{Jiajun Liao}
\emailAdd{liaojiajun@mail.sysu.edu.cn}
\affiliation[b]{
School of Physics, Sun Yat-Sen University, Guangzhou 510275, China}

\abstract{
A connection between the neutrino and an exotic fermion is described in the general neutrino model. In this model the neutrinos can convert into the new fermion and thus the interaction leads to novel recoil spectrum in the neutrino scattering experiments.
We study the general neutrino interaction by evaluating both the tree-level and loop-level contributions to the coherent elastic neutrino-nucleus scattering. We illustrate the scattering by taking the framework of a simplified neutrino model with a Dirac fermion $\chi$ and a spin-0 mediator. For the CP phase in the quark sector being 0 and $\pi/2$, the detection processes are dominated by the tree-level and loop-level contribution, respectively. We investigate the constraints on the couplings between the mediator and the new particle $\chi$ or the quarks by fitting to the COHERENT data. The parameter space with $m_\chi$ larger than the maximal energy of incoming neutrinos can be also constrained by including the loop-level contribution.
}

\arxivnumber{arXiv:2008.00743}

\begin{document}

\maketitle
\setcounter{page}{2}

\newpage

\section{Introduction}
\label{sec:Intro}

The COHERENT experiment has observed the coherent elastic neutrino-nucleus scattering (CE$\nu$NS) process at the $6.7\sigma$
level~\cite{Akimov:2017ade}. The neutrinos measured at COHERENT are produced at the Spallation Neutron Source (SNS) by stopped pion and muon decays with energies $E\lesssim 53$ MeV. The CE$\nu$NS process occurs when the moment transfer in the neutrino-nucleus scattering process is smaller than the inverse of the target nucleus radius, and the scattering amplitudes of the nucleons inside the nucleus can be added coherently, which leads to a large enhancement of the cross section.
The observation of CE$\nu$NS at COHERENT is consistent with the prediction of the Standard Model (SM), in which the CE$\nu$NS
process is generated through the weak neutral current~\cite{Freedman:1973yd}. Besides the
active neutrinos through $Z$ boson exchange in the SM, any neutrino
flavors including light right-handed (RH) neutrinos can be produced in the final state of the CE$\nu$NS process. Here the RH neutrinos refer to sterile neutrinos which do not carry any SM gauge charges. Since the production of the RH neutrinos will not violate the coherence condition if the momentum transfer are smaller than the inverse of the nucleus radius, the COHERENT observation thus provides us an opportunity to
explore the new physics (NP) associated with general neutrino interactions in the presence of
exotic fermion such as the RH neutrino.

Recently, different groups studied the conversion to an exotic fermion which could or could not be a dark matter (DM) particle~\footnote{Note that the inverse process in which the exotic fermion as DM particle can be absorbed by the target and emit a neutrino could lead to distinct DM signal in DM direct detection. However, the decaying DM scenario usually faces the requirement of stability. Requiring
the DM being stable at the Universe time scale would set a very stringent bound on the coupling
and/or the DM mass. We will not interpret $\chi$ as a DM candidate and discuss the reverse process to detect this kind of DM scenario. We refer the discussion of the relevant neutrino-portal DM to Refs.~\cite{Brdar:2018qqj,Dror:2019onn,Dror:2019dib,Hurtado:2020vlj,Dror:2020czw}.} in the coherent elastic neutrino-nucleus scattering~\cite{Farzan:2018gtr,Brdar:2018qqj,Chang:2020jwl,Hurtado:2020vlj}. The calculation of CE$\nu$NS process depends on the specific interactions between neutrino and SM quark sector.
For instance, in the studies of neutrino interaction in Ref.~\cite{Hurtado:2020vlj}, the authors assumed that the interaction is mediated by a scalar field $a$ via the Yukawa couplings $g_\chi \bar{\chi}\nu a + g_q\bar{q}q a$. By contrast, if we consider the mediator $a$ as a pseudoscalar, the pseudoscalar quark current $\bar{q}i\gamma_5 q a$ would lead to effective coupling for nuclear spin-dependent (SD) interaction which is determined by a sum over spin-up and spin-down nucleons with opposite signs~\cite{Freedman:1977xn}. Thus, this kind of contribution to CE$\nu$NS is highly suppressed and usually neglected in the analysis of general neutrino interactions for heavy CsI nuclei in COHERENT experiment~\cite{Akimov:2017ade}.
However, it is worth emphasizing that the tree-level interactions induced by pseudoscalar quark current can generate loop diagrams which in turn give scalar interactions and non-momentum-suppressed spin-independent (SI) scattering. The nuclear matrix element from these SI interactions could receive the enhancement of nuclear mass number in coherent scattering and compensate the suppression from the perturbative loop calculation. As a result, the full calculation involving the loop corrections leads to detectable observation for the pseudoscalar interaction.
This loop effect has been taken into account in both simplified frameworks and UV complete models for detecting the
Weakly Interacting Massive Particle (WIMP) in direct DM detection~\cite{Drees:1993bu,Hisano:2010ct,Baek:2016lnv,Baek:2017ykw,Arcadi:2017wqi,Li:2018qip,Abe:2018emu,Abe:2018bpo,Li:2019fnn,Mohan:2019zrk,
Ertas:2019dew,Giacchino:2015hvk,Giacchino:2014moa,Ibarra:2014qma,Colucci:2018vxz,Colucci:2018qml,Chao:2019lhb}.

In this work we investigate both the tree-level and loop-level contributions to the CE$\nu$NS in the framework of a simplified neutrino model. We assume generic neutrino currents interacting with an exotic fermion $\chi$ and SM quarks through a light spin-0 mediator $a$ with general CP phases. For the CE$\nu$NS process, besides the scattering $\nu N\to \chi N$ with neutrinos converting to $\chi$ at tree-level, the loop diagrams can also induce elastic scattering $\nu N\to \nu N$ with the intermediate new fermion $\chi$ inside the loops. This additional contribution will affect the fit to COHERENT data and play an important role when the scattering process $\nu N\to \chi N$ is kinematically forbidden or the scalar current $\bar{q}q a$ is absent.

This paper is organized as follows. In Sec.~\ref{sec:Model} we describe the simplified neutrino model. In Sec.~\ref{sec:CEnuNT} we present the analytical expressions of the CE$\nu$NS cross section. Both the tree-level and loop-level contributions are given in general forms. The numerical results are also shown. We discuss other relevant constraints on this model in Sec.~\ref{sec:Cons}. Our conclusions are drawn in Sec.~\ref{sec:Con}. Some calculational details are collected in the Appendix.

\section{Simplified neutrino model with an exotic fermion}
\label{sec:Model}

We consider a Dirac fermion $\chi$ charged under lepton number being a SM gauge singlet. It is generally viewed as the sterile neutrino but can be a generic singlet fermion which mixes with the neutrino fields $\nu$. In the simplified neutrino model, the neutrinos interact with the new fermion $\chi$ through a spin-0 field $a$ and the mediator $a$ couples to the SM quarks as
\begin{eqnarray}
\mathcal{L}\supset g_\chi a \bar{\chi}\Big(\cos\theta_\chi + i\gamma_5\sin\theta_\chi \Big)\nu + \sum_q g_q a \bar{q}\Big(\cos\theta_q + i\gamma_5 \sin\theta_q \Big) q  + h.c.\; ,
\label{eq:Lagrangian}
\end{eqnarray}
where $\theta_\chi$ denotes the relative CP phase angle between $\nu$ and $\chi$,
and $\theta_q$ is the CP phase in quark sector.
Here we use a universal coupling $g_q$ for the interaction between SM quarks and the mediator $a$. In practice, below we perform a generic hypothesis with different $g_q$ couplings for up-type and down-type quarks. This hypothesis can be realized in the two-Higgs-doublet model (2HDM)~\cite{He:2008qm,Abe:2018emu,Abe:2018bpo} and one will see that the flavor constraints can be relaxed in this choice.
The choice of $(\theta_\chi,\theta_q)=(0,0)$ is exactly the case with pure scalar mediator studied in Ref.~\cite{Hurtado:2020vlj}. The other CP conserving case is $(\theta_\chi,\theta_q)=(\pi/2,\pi/2)$ with the field $a$ being a pseudoscalar mediator. Generally, the scenario with $\theta_q=\pi/2$ leads to suppressed SD interaction at tree-level and the detectable signals are absent in neutrino experiments.

One should note that the above simplified hypothesis does not respect gauge invariance prior to the SM electroweak symmetry breaking. Thus, we expect that there exist additional couplings between $a$ and the SM Higgs in specific UV complete models~\cite{Abe:2018emu}. Without loss of generality, we introduce a scalar trilinear coupling
\begin{eqnarray}
\mathcal{L}\supset {1\over 2}\lambda_{haa} v_0 h a^2\;,
\label{eq:trilinear}
\end{eqnarray}
where $v_0=(\sqrt{2}G_F)^{-1/2}\simeq 246$ GeV is the SM Higgs vacuum expectation value (vev).
We will show that this interaction induces additional loop diagram in the scattering process.

\begin{figure}[h!]
\begin{center}
\minigraph{5cm}{-0.05in}{(a)}{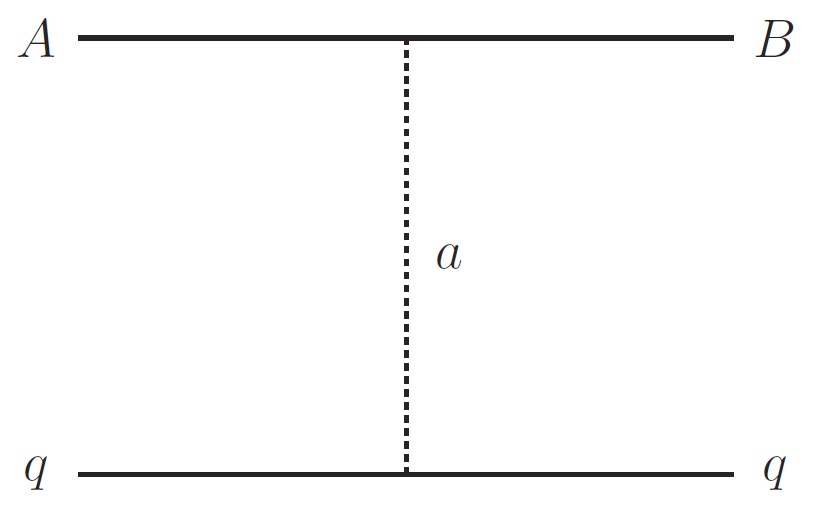}
\minigraph{5cm}{-0.05in}{(b)}{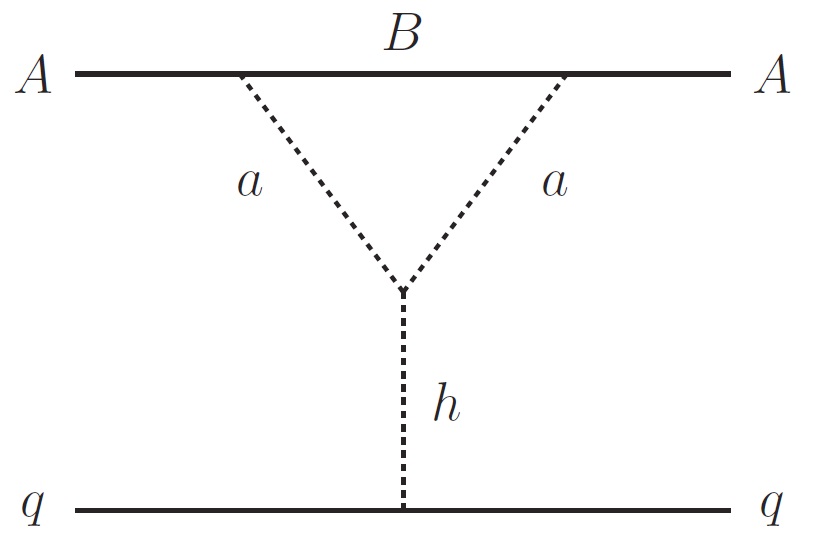}\\
\minigraph{5cm}{-0.05in}{(c)}{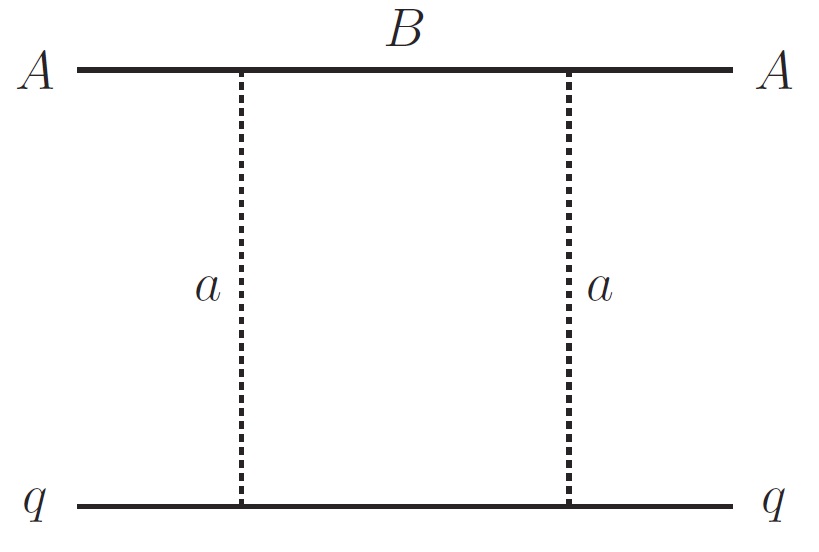}
\minigraph{5cm}{-0.05in}{(d)}{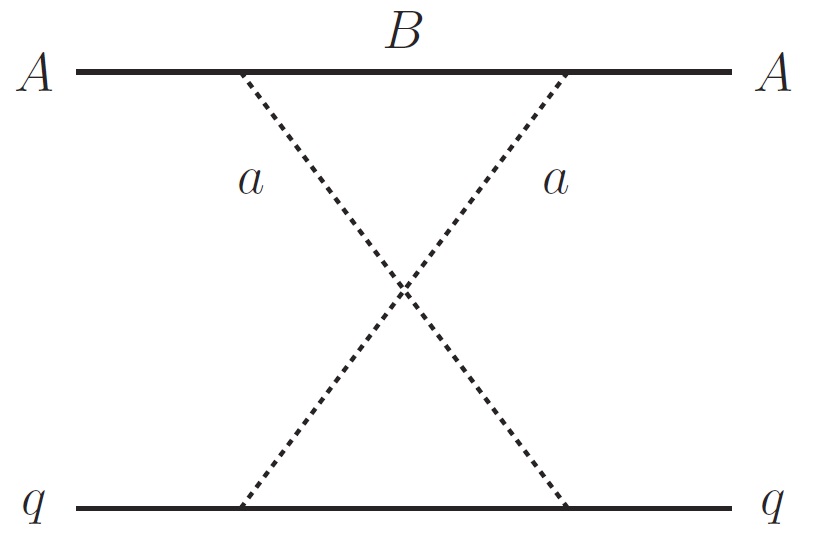}\\
\minigraph{5cm}{-0.05in}{(e)}{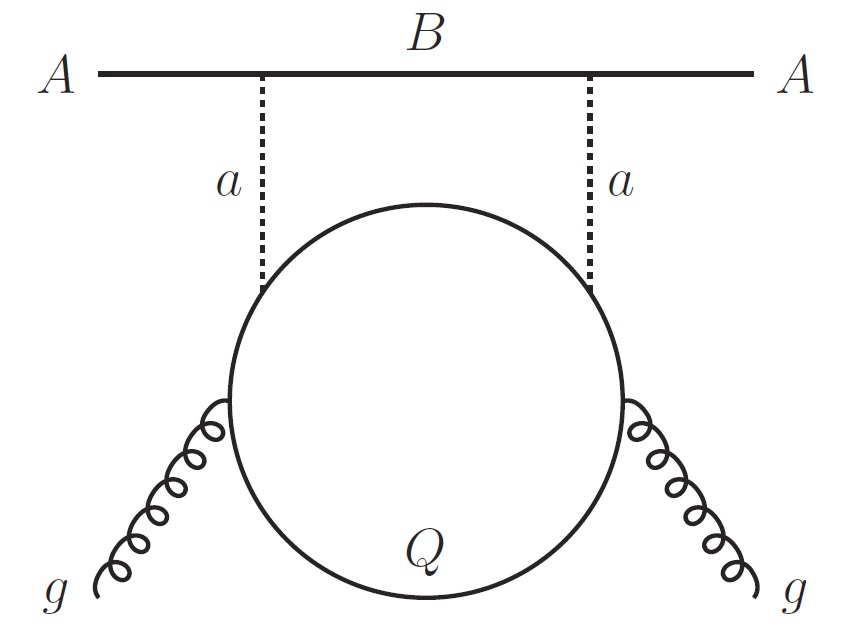}
\minigraph{5cm}{-0.05in}{(f)}{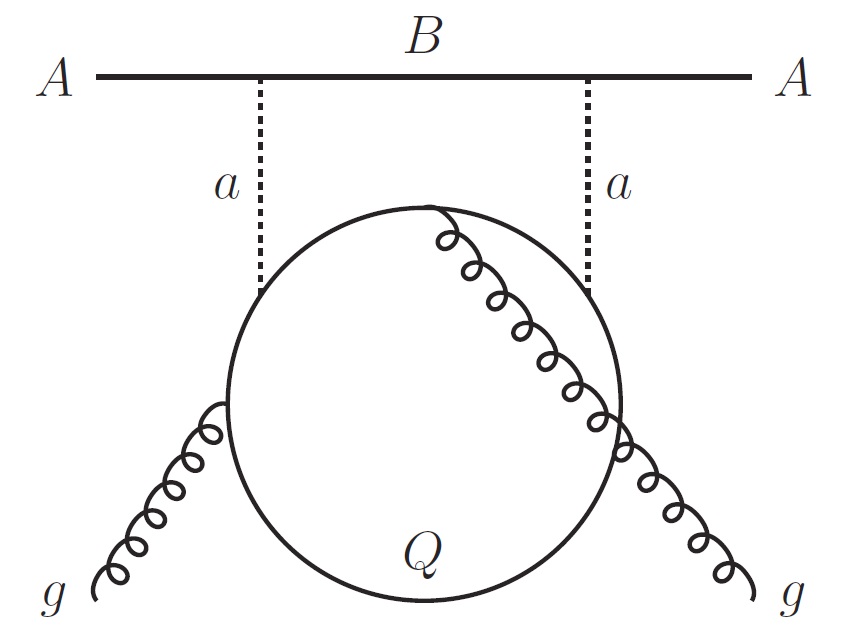}
\end{center}
\caption{Diagrams for the processes of neutrino interacting with quark or gluon. For incoming neutrino processes, $A=\nu, B=\chi$. }
\label{diagram}
\end{figure}

In Fig.~\ref{diagram} we show the diagrams for the processes of neutrino interacting with the nucleus target at quark-level.
For incoming neutrino processes, $A=\nu, B=\chi$. The tree-level diagram in Fig.~\ref{diagram} (a) is generated by two vertexes from
the two couplings in the Lagrangian given in Eq.~(\ref{eq:Lagrangian}) and a mediator $a$ in t channel. Fig.~\ref{diagram} (b) is a triangle diagram intermediated by the $\lambda_{haa}$ coupling and the SM Higgs field $h$. Figs.~\ref{diagram} (c) and (d) are the box diagrams formed by two $a$ fields in the internal lines. Recent developments in the loop calculation include the contributions from
two-loop scattering diagrams in Figs.~\ref{diagram} (e) and (f) for scalar-type gluon operator ${\alpha_s\over \pi} G^a_{\mu\nu}G^{a\mu\nu}$~\cite{Abe:2018emu,Li:2019fnn,Ertas:2019dew}. The full two-loop calculation can be obtained by integrating out the heavy quarks and the mediators in the loops. In next sections we will display how the effective operators for the neutrino scattering are formed from these diagrams.

\section{Coherent elastic neutrino-nucleus scattering}
\label{sec:CEnuNT}

For the incoming neutrino scattering $\nu(p_1) q(k_1)\to \chi(p_2) q(k_2)$ in Fig.~\ref{diagram} (a) with $A=\nu, B=\chi$, we obtain the tree-level matrix element
\begin{eqnarray}
iM_{\rm tree}&=&\sum_{q=all} {-i\over t-m_a^2}g_\chi g_q \bar{\chi}(p_2) (c_{\theta_\chi}+i\gamma_5 s_{\theta_\chi}) P_L \nu(p_1)[c_{\theta_q}\bar{q}(k_2)q(k_1)+s_{\theta_q}\bar{q}(k_2)i\gamma_5 q(k_1)]\;,
\end{eqnarray}
where $P_L={1-\gamma_5\over 2}$, the Mandelstam variable $t=(p_1-p_2)^2$ and $c_x\equiv \cos(x), s_x\equiv \sin(x)$. When $(\theta_\chi,\theta_q)=(0,0)$, this is exactly the case in Ref.~\cite{Hurtado:2020vlj} with scalar mediator. In the case of $\theta_q=\pi/2$, the SI contribution from tree-level diagram to the neutrino-nucleus scattering is zero.

For the elastic scattering process $\nu(p_1) q(k_1)\to \nu(p_2) q(k_2)$, there are multiple loop-level contributions as shown in Fig.~\ref{diagram}. We first consider the one-loop triangle diagram with SM Higgs exchange and the matrix element is
\begin{eqnarray}
iM_{\rm triangle}
&=&\sum_{q=all} {-i\lambda_{haa}v_0 m_\chi g_\chi^2\over (4\pi)^2(t-m_h^2)} C_0[p_1^2, (p_1-p_2)^2, p_2^2; m_\chi^2, m_a^2, m_a^2]\nonumber \\
&\times&\bar{\nu}(p_2) (c_{2\theta_\chi}+i\gamma_5 s_{2\theta_\chi}) P_L \nu(p_1) \bar{q}(k_2)q(k_1)\;.
\end{eqnarray}
In the zero momentum transfer limit $t\to 0$ and taking massless neutrinos, the Passarino-Veltman function $C_0$ can be simplified as shown in the Appendix.
One can see that the quark current is pure scalar-type in this diagram due to the SM Higgs exchange.
The matrix element of the one-loop box diagrams is
\begin{eqnarray}
iM_{\rm box}&=&\sum_{q=u,d,s}{i\over (4\pi)^2} g_\chi^2 g_q^2  {4m_\chi m_q\over m_a^2}D_{00} \bar{\nu}(p_2)(c_{2\theta_\chi}+i\gamma_5 s_{2\theta_\chi})P_L\nu(p_1)\bar{q}(k_2)q(k_1)\nonumber \\
&+&\sum_{q=u,d,s}{i\over (4\pi)^2} g_\chi^2 g_q^2  4m_\chi m_q c_{\theta_q}D_{0} \bar{\nu}(p_2)(c_{2\theta_\chi}+i\gamma_5 s_{2\theta_\chi})P_L\nu(p_1)\bar{q}(k_2)(c_{\theta_q}+i\gamma_5s_{\theta_q})q(k_1)\nonumber \\
&+&\sum_{q=u,d,s,c,b}{i\over (4\pi)^2} g_\chi^2 g_q^2 {8\over m_a^2}D_{001}\bar{\nu}(p_2)i\partial^\mu \gamma^\nu P_L\nu(p_1) O_{\mu\nu}^q\nonumber \\
&+&\sum_{q=u,d,s,c,b}{i\over (4\pi)^2} g_\chi^2 g_q^2 {4m_\chi\over m_a^2}D_{11} \bar{\nu}(p_2)(c_{2\theta_\chi}+i\gamma_5 s_{2\theta_\chi})i\partial^\mu i\partial^\nu P_L\nu(p_1)O_{\mu\nu}^q\;,
\end{eqnarray}
where $O_{\mu\nu}^q$ is the twist-2 operator for quark
\begin{eqnarray}
O_{\mu\nu}^q={i\over 2}\bar{q}\Big(\partial_\mu\gamma_\nu+\partial_\nu\gamma_\mu-{1\over 2}g_{\mu\nu}\cancel{\partial}\Big)q\;.
\end{eqnarray}
The Passarino-Veltman functions are also collected in the Appendix.

For the heavy quark loops in the two-loop diagrams, we calculate the amplitude using the Fock-Schwinger gauge for the gluon background field~\cite{Novikov:1983gd,Hisano:2010ct}. First, the amplitude contributing to the effective operator $aaG^a_{\mu\nu}G^{a\mu\nu}$ is
\begin{eqnarray}
iM_{aaGG}=i\Pi_{G}(\ell^2){\alpha_s\over 12\pi}G^a_{\rho\sigma}G^{a\rho\sigma}+i\Pi_{\tilde{G}}(\ell^2){\alpha_s\over 8\pi}G^a_{\rho\sigma}\tilde{G}^{a\rho\sigma}\;,
\label{ampaaGG}
\end{eqnarray}
where $G^{a\mu\nu}$ is the gluon field strength tensor and $\tilde{G}^{a\mu\nu}={1\over 2}\epsilon^{\mu\nu\alpha\beta}G^a_{\alpha\beta}$.
Then, the complete two-loop matrix element in Figs.~\ref{diagram} (e) and (f) reads
\begin{eqnarray}
iM_{\rm 2-loop}&=& -g_\chi^2 \int {d^4\ell\over (2\pi)^4} \bar{\nu}(p_2)[\cancel{\ell}+m_\chi(c_{2\theta_\chi}+i\gamma_5 s_{2\theta_\chi})]P_L \nu(p_1)\nonumber \\
&\times& {1\over [(\ell+p_1)^2-m_\chi^2](\ell^2-m_a^2)^2}\Big[\Pi_G(\ell^2){\alpha_s\over 12\pi}G^a_{\rho\sigma}G^{a\rho\sigma}+ \Pi_{\tilde{G}}(\ell^2){\alpha_s\over 8\pi}G^a_{\rho\sigma}\tilde{G}^{a\rho\sigma}\Big]\nonumber \\
&=&\Big[C_{G,S}\bar{\nu}(p_2)P_L\nu(p_1)+C_{G,PS}\bar{\nu}(p_2)i\gamma_5 P_L\nu(p_1)\Big]{-\alpha_s\over 12\pi}G^a_{\rho\sigma}G^{a\rho\sigma}\nonumber \\
&+&\Big[C_{\tilde{G},S}\bar{\nu}(p_2)P_L\nu(p_1)+C_{\tilde{G},PS}\bar{\nu}(p_2)i\gamma_5 P_L\nu(p_1)\Big]{\alpha_s\over 8\pi}G^a_{\rho\sigma}\tilde{G}^{a\rho\sigma} \;,
\end{eqnarray}
where $\ell$ denotes the momentum of the mediator $a$ and
\begin{eqnarray}
C_{G,S}&=&{i\over (4\pi)^2} \sum_{Q=c,b,t}g_\chi^2 g_Q^2 m_\chi c_{2\theta_\chi} F_{G}(p_1^2,m_\chi^2,m_a^2,m_Q^2) \;,\\
C_{G,PS}&=&{i\over (4\pi)^2} \sum_{Q=c,b,t}g_\chi^2 g_Q^2 m_\chi s_{2\theta_\chi} F_{G}(p_1^2,m_\chi^2,m_a^2,m_Q^2) \;,\\
C_{\tilde{G},S}&=&-{i\over (4\pi)^2} \sum_{Q=c,b,t}g_\chi^2 g_Q^2 m_\chi c_{2\theta_\chi} F_{\tilde{G}}(p_1^2,m_\chi^2,m_a^2,m_Q^2) \;,\\
C_{\tilde{G},PS}&=&-{i\over (4\pi)^2} \sum_{Q=c,b,t}g_\chi^2 g_Q^2 m_\chi s_{2\theta_\chi} F_{\tilde{G}}(p_1^2,m_\chi^2,m_a^2,m_Q^2) \;.
\end{eqnarray}
The above $\Pi_{G}(\ell^2), \Pi_{\tilde{G}}(\ell^2)$ and $F_G, F_{\tilde{G}}$ functions are all given in the Appendix.

The nucleon form factors are defined as~\cite{DelNobile:2013sia,Bishara:2017pfq}
\begin{eqnarray}
\langle N| m_q \bar{q}q|N\rangle&=&m_N f_q^N \bar{N}N\;, \quad q=u,d,s\;,\\
\langle N| m_Q \bar{Q}Q|N\rangle&=&\langle N| {-\alpha_s\over 12\pi} G^a_{\mu\nu}G^{a\mu\nu}|N\rangle = {2\over 27} m_N f_G^N \bar{N}N\;, \quad Q=c,b,t\;,\\
\langle N| O_{\mu\nu}^q|N\rangle&=& {1\over m_N} \Big(p^N_\mu p^N_\nu -{1\over 4}m_N^2 g_{\mu\nu} \Big)\Big(q^N(2)+\bar{q}^N(2)\Big) \bar{N}N\;, \quad q=u,d,s,c,b\;,
\end{eqnarray}
for the SI interactions and those for SD interactions are
\begin{eqnarray}
\langle N| m_q \bar{q}i\gamma_5 q|N\rangle&=& F_P^{q/N}(q^2) \bar{N}i\gamma_5 N\;, \quad q=u,d,s\;,\\
\langle N| m_Q \bar{Q}i\gamma_5 Q|N\rangle&=&\langle N| {\alpha_s\over 8\pi} G^a_{\mu\nu}\tilde{G}^{a\mu\nu}|N\rangle = F_{\tilde{G}}^N(q^2) \bar{N}i\gamma_5 N\;, \quad Q=c,b,t\;.
\end{eqnarray}
Next, we can obtain the matrix elements at nucleon-level
\begin{eqnarray}
iM_{\rm tree}^N
&=& {-i\over t-m_a^2}g_\chi g_q c_{\theta_q} \Big(\sum_{q=u,d,s} {m_N\over m_q} f_q^N+\sum_{Q=c,b,t}{2\over 27}{m_N\over m_Q}f_G^N\Big)\nonumber \\
&\times&\bar{\chi}(p_2) (c_{\theta_\chi}+i\gamma_5 s_{\theta_\chi}) P_L \nu(p_1)\bar{N}(k_2)N(k_1)+\fbox{SD}\;,
\label{eq:treeM}
\end{eqnarray}
\begin{eqnarray}
iM_{\rm triangle}^N&=& {-i\lambda_{haa}v_0 m_\chi g_\chi^2\over (4\pi)^2(t-m_h^2)} C_0 \Big(\sum_{q=u,d,s} {m_N\over m_q} f_q^N+\sum_{Q=c,b,t}{2\over 27}{m_N\over m_Q}f_G^N\Big)\nonumber \\
&\times&\bar{\nu}(p_2) (c_{2\theta_\chi}+i\gamma_5 s_{2\theta_\chi}) P_L \nu(p_1) \bar{N}(k_2)N(k_1)\;,
\label{eq:triangleM}
\end{eqnarray}
\begin{eqnarray}
iM_{\rm box}^N&=&
\sum_{q=u,d,s}{i\over (4\pi)^2} g_\chi^2 g_q^2 f_q^N \Big({4m_\chi m_N\over m_a^2}D_{00}+4m_\chi m_N c_{\theta_q}^2 D_{0}\Big) \nonumber \\
&\times&\bar{\nu}(p_2)(c_{2\theta_\chi}+i\gamma_5 s_{2\theta_\chi})P_L\nu(p_1)\bar{N}(k_2)N(k_1)+\fbox{SD}\;,
\label{eq:boxM}
\end{eqnarray}
\begin{eqnarray}
iM_{\rm 2-loop}^N&=&\Big[C_{G,S}\bar{\nu}(p_2)P_L\nu(p_1)+C_{G,PS}\bar{\nu}(p_2)i\gamma_5 P_L\nu(p_1)\Big]{2\over 27} m_N f_G^N \bar{N}(k_2)N(k_1)+\fbox{SD} \;.
\label{eq:2loopM}
\end{eqnarray}
Here $\fbox{SD}$ stands for SD terms which will be omitted in the following calculation.

Since $\chi$ is not detected in a neutrino scattering experiment, the total differential cross section of CE$\nu$NS can be written as
\begin{align}
\frac{d\sigma}{dT}=\frac{d\sigma_\text{SM}}{dT}+\frac{d\sigma_\text{tree}}{dT}+\frac{d\sigma_\text{loop}}{dT}\,,
\end{align}
where $T$ is the nuclear recoil energy.
The SM differential cross section is given by
\begin{align}
\frac{d\sigma_\text{SM}}{dT}=\frac{G_F^2 M}{2\pi}[Z g_p^V + N g_n^V]^2F^2(Q^2)(2-\frac{MT}{E^2})\,,
\end{align}
where $F(Q^2)$ refers to the nuclear form factor with the moment transfer $Q^2=2MT$. We take the Helm parameterization~\cite{Helm:1956zz} for the nuclear form factor. Note that employing a different form factor parameterization has a negligible effect on the COHERENT spectrum~\cite{Cadeddu:2017etk, AristizabalSierra:2019zmy}, and the form factor uncertainty driven by the nucleon density distribution has been taken into account in our analysis. Here $M$ is the mass of target nucleus, $E$ is the incoming neutrino energy,  $Z$ ($N$) is the number of protons (neutrons) in the target nucleus,  $g_n^V=-\frac{1}{2}$ and $g_p^V=\frac{1}{2}-2\sin^2\theta_W$ are the SM weak couplings with $\theta_W$ being the weak mixing angle.

From Eq.~(\ref{eq:treeM}), the tree-level differential cross section of $\nu N\to \chi N$ is
\begin{align}
\frac{d\sigma_\text{tree}}{dT}=\frac{g_\chi^2 g_N^2F^2(Q^2)}{16\pi E^2 (m_a^2+2MT)^2}\left(2 M+T\right)\left( 2MT+ m_\chi^2 \right)\,,
\end{align}
where
\begin{align}
g_N=Z m_p \left(\sum_{q=u,d,s} {f_q^p\over m_q}+\sum_{Q=c,b,t}{2\over 27}{f_G^p\over m_Q}\right)g_qc_{\theta_q}+N m_n \left(\sum_{q=u,d,s} {f_q^n\over m_q}+\sum_{Q=c,b,t}{2\over 27}{f_G^n\over m_Q}\right)g_qc_{\theta_q}\,.
\label{eq:treelevel}
\end{align}
Note that in order to produce a massive fermion $\chi$ in the scattering $\nu N\to \chi N$, the energy of the incident neutrinos should be larger than a minimal energy~\cite{Brdar:2018qqj, Chang:2020jwl}, i.e.,
\begin{align}
E>m_\chi+\frac{m_\chi^2}{2M}\,.
\end{align}

From Eqs.~(\ref{eq:triangleM}),~(\ref{eq:boxM}) and~(\ref{eq:2loopM}),
we can write the loop-level differential cross section as
\begin{align}
\frac{d\sigma_\text{loop}}{dT}=\frac{G_{\rm loop}^2M}{8\pi E^2}F^2(Q^2)\left( 2MT+ T^2 \right)\,,
\end{align}
where
\begin{align}
&G_{\rm loop}={Z\over (4\pi)^2}g_\chi^2m_\chi m_p\left[\frac{\lambda_{haa} v_0}{2MT+m_h^2}\left(\sum_{q=u,d,s} {f_q^p\over m_q}+\sum_{Q=c,b,t}{2\over 27}{f_G^p\over m_Q}\right)C_0\right.
\\\nonumber
&+\left.4\sum_{q=u,d,s}g_q^2 f_q^p \Big({D_{00}\over m_a^2}+ c_{\theta_q}^2 D_{0}\Big)+{2\over 27} f_G^p\sum_{Q=c,b,t} g_Q^2 F_{G}(p_1^2,m_\chi^2,m_a^2,m_Q^2)\right]
\\\nonumber
&+{N\over (4\pi)^2}g_\chi^2m_\chi m_n\left[\frac{\lambda_{haa}v_0}{2MT+m_h^2}\left(\sum_{q=u,d,s} {f_q^n\over m_q}+\sum_{Q=c,b,t}{2\over 27}{f_G^n\over m_Q}\right)C_0\right.
\\\nonumber
&+\left.4\sum_{q=u,d,s}g_q^2 f_q^n \Big({D_{00}\over m_a^2}+ c_{\theta_q}^2 D_{0}\Big)+{2\over 27} f_G^n\sum_{Q=c,b,t} g_Q^2 F_{G}(p_1^2,m_\chi^2,m_a^2,m_Q^2)\right]\,.
\end{align}
One can see that the total differential cross section has no dependence on the mixing angle $\theta_\chi$. This is because the tree-level amplitude and the loop-level amplitude have no interference in this case and the neutrinos in the external legs are nearly massless.
The measurement of the SM Higgs decay at the LHC~\cite{Khachatryan:2016vau} implies the constraint on the coupling $\lambda_{haa}\lesssim 0.01$~\cite{Ertas:2019dew}. In addition, the triangle diagram is suppressed by a factor of $m_\chi v_0/m_h^2\sim \mathcal{O}(10^{-5})$ for $m_\chi \sim \mathcal{O}(1)$ MeV. The flavor physics also sets stringent constraints on the coupling $g_q$ for up-type quarks as discussed below. Thus, in the numerical calculation, we neglect the coupling for up-type quarks and the triangle diagram and the two-loop diagrams dominated by top quark contribution. We checked that this ignorance does not affect our conclusion. For the nucleon form factors in SI interactions, we adopt the default values in micrOMEGAs~\cite{Belanger:2008sj,Belanger:2018ccd}.

The CE$\nu$NS process has been recently observed by the COHERENT experiment in a low-threshold CsI detector at the 6.7$\sigma$ CL. The neutrinos measured at COHERENT are generated from the stopped pion decays and the muon decays, and their fluxes are well known and given by
\begin{align}
\label{eq:nu-spectra.COHERENT}
\phi_{\nu_\mu}(E_{\nu_\mu})&= \mathcal{N}_0
\frac{2m_\pi}{m_\pi^2-m_\mu^2}\,
\delta\left(
1-\frac{2E_{\nu_\mu}m_\pi}{m_\pi^2-m_\mu^2}
\right) \ ,
\nonumber\\
\phi_{\nu_e}(E_{\nu_e})&= \mathcal{N}_0 \frac{192}{m_\mu}
\left(\frac{E_{\nu_e}}{m_\mu}\right)^2
\left(\frac{1}{2}-\frac{E_{\nu_e}}{m_\mu}\right)\ ,\\
\phi_{\overline\nu_\mu}(E_{\overline\nu_\mu})&= \mathcal{N}_0  \frac{64}{m_\mu}
\left(\frac{E_{\overline\nu_\mu}}{m_\mu}\right)^2
\left(\frac{3}{4}-\frac{E_{\overline\nu_\mu}}{m_\mu}\right)\,,\nonumber
\end{align}
where $\mathcal{N}_0$ is a normalization factor determined by the setup of the COHERENT experiment. The $\nu_\mu$ component is produced from the stopped pion decays, $\pi^+\to \mu^++\nu_\mu$, which yield a monoenergetic flux at $(m^2_{\pi}-m_\mu^2)/(2 m_\pi)\simeq 30$~MeV. The $\bar{\nu}_\mu$ and $\nu_e$ components are produced from the subsequent muon decays, $\mu^+\to e^++\bar{\nu}_\mu+\nu_e$, and their energies have a kinematic upper bound at $m_\mu/2 \simeq 53$ MeV.

\begin{figure}[htb!]
\begin{center}
\includegraphics[scale=1,width=0.65\linewidth]{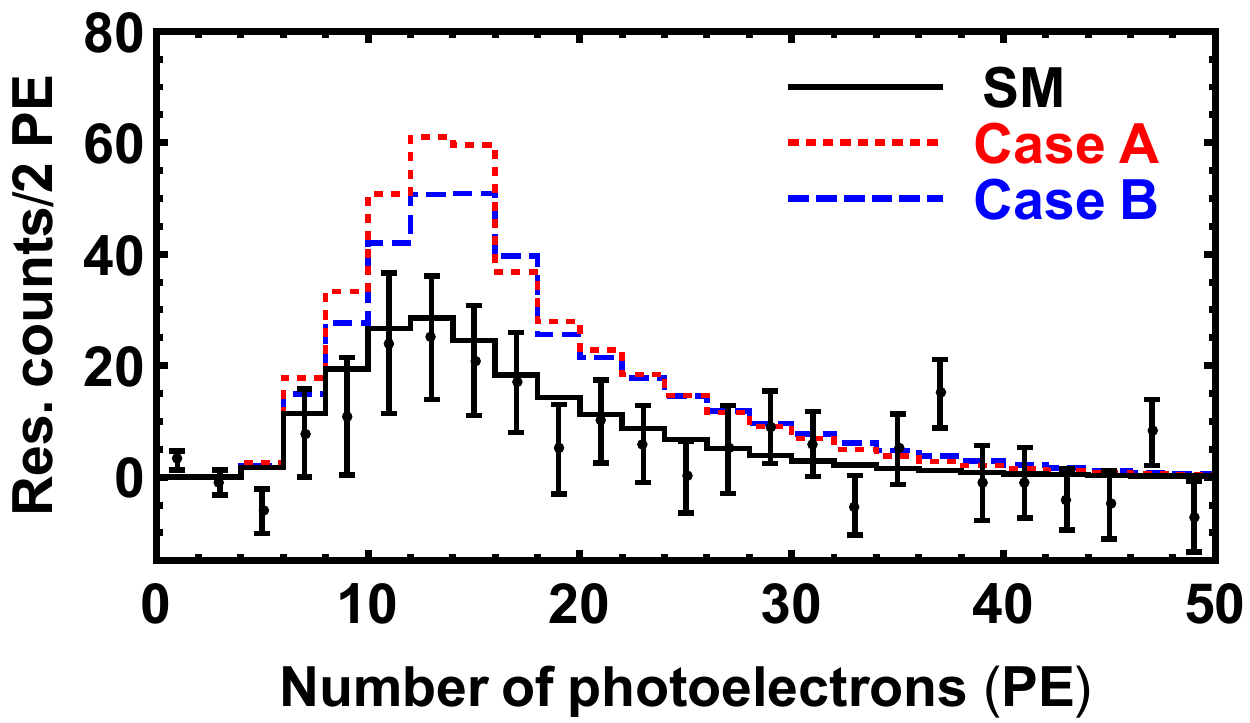}
\end{center}
\caption{
The expected CE$\nu$NS residual event as a function of the number of photoelectrons at COHERENT. The black solid lines correspond to the SM case, and the red dotted (blue dashed) lines correspond to Case A (B) with  $\theta_q=0$, $m_\chi=10$ MeV, $m_a=100$ MeV, and $g_\chi g_q=5.0\times10^{-9}$ ($\theta_q=\pi/2$, $m_\chi=100$ MeV, $m_a=200$ MeV, and $g_\chi g_q=0.01$).
}
\label{fig:spectrum}
\end{figure}

The presence of $\chi$-neutrino interaction will modify the COHERENT spectrum, which can be seen in Fig.~\ref{fig:spectrum}. We select two benchmark points to illustrate the effects of modified spectra:
\begin{itemize}
	\item Case A: $\theta_q=0$, $m_\chi=10$ MeV, $m_a=100$ MeV, and $g_\chi g_q=5.0\times10^{-9}$\;,
	\item Case B: $\theta_q=\pi/2$, $m_\chi=100$ MeV, $m_a=200$ MeV, and $g_\chi g_q=0.01$\;.
\end{itemize}
In Case A, the modification to the SM spectrum is dominated by the tree-level scattering process, $\nu N\to \chi N$, and the loop-level contribution is negligible due to the small coupling constants. In Case B, since $m_\chi\gtrsim 53$ MeV, the tree-level process is kinematically forbidden, and the modification to the SM spectrum is only contributed by the loop-level diagrams.

Following Ref.~\cite{Chang:2020jwl}, we evaluate the statistical significance of BSM by defining
\begin{align}
\chi^2 = \sum_{i=4}^{15} \left[\frac{N_\text{meas}^i-N_\text{th}^i(1+\alpha)-B_\text{on}(1+\beta)}{\sigma_\text{stat}^i}\right]^2+\left(\frac{\alpha}{\sigma_\alpha}\right)^2+\left(\frac{\beta}{\sigma_\beta}\right)^2\,,
\end{align}
where $N_\text{meas}^i$ and $N_\text{th}^i$ denotes the number of measured (predicted) events per energy bin, respectively. $\alpha$ ($\beta$) represents the nuisance parameters for the signal rate (the beam-on background) with a uncertainty of $\sigma_\alpha=0.28$ ($\sigma_\beta=0.25$)~\cite{Akimov:2017ade}. The neutrino flux uncertainty (10\%), form factor uncertainty (5\%), signal acceptance uncertainty (5\%), and quenching factor uncertainty (25\%) are included in the signal rate uncertainty $\sigma_\alpha$.
The statistical uncertainty per energy bin is calculated by $\sigma_\text{stat}^i=\sqrt{ N_\text{meas}^i +2B_\text{SS}^i+B_\text{on}^i}$ with $B_\text{SS}$ being the steady-state background from the anti-coincident data, and $B_\text{on}$ the beam-on background mainly consists of prompt neutrons.

\begin{figure}[htb!]
\begin{center}
\includegraphics[scale=1,width=0.48\linewidth]{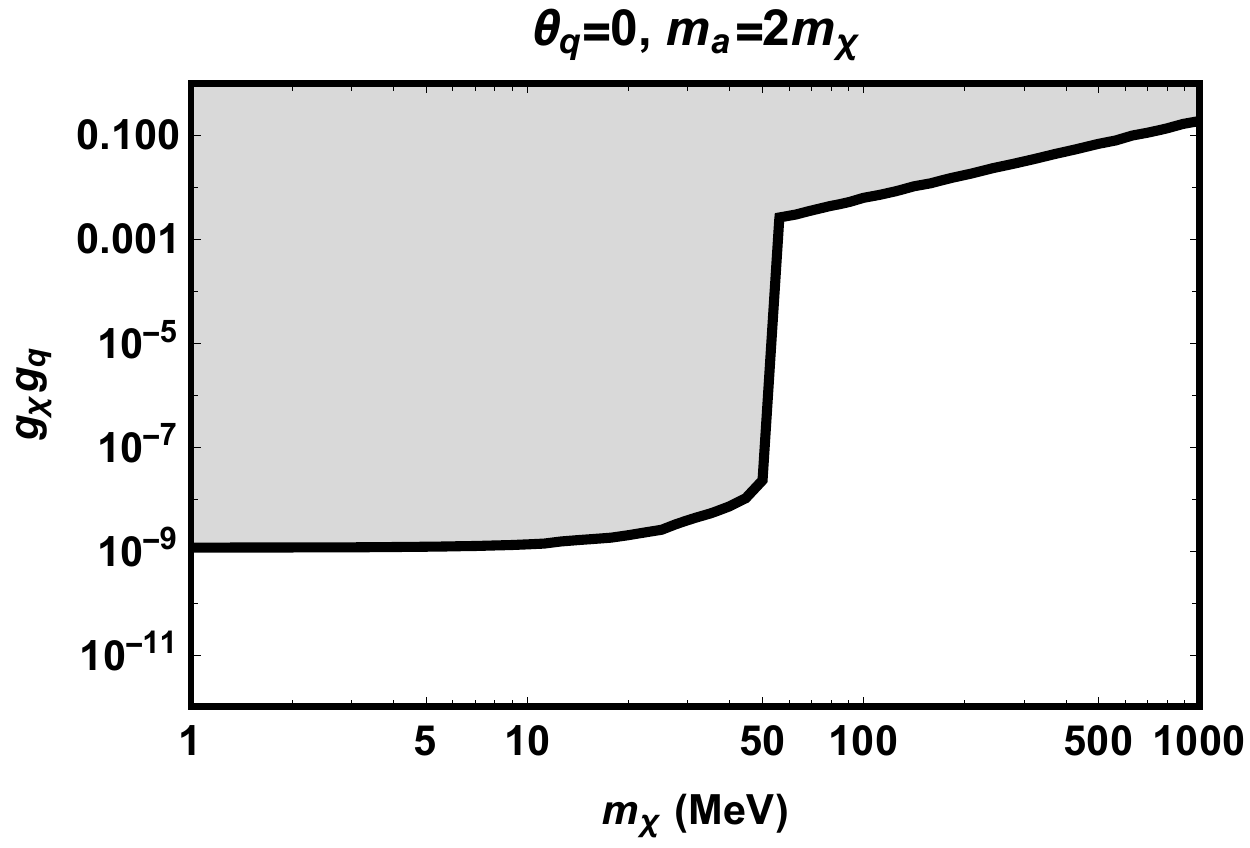}
\includegraphics[scale=1,width=0.48\linewidth]{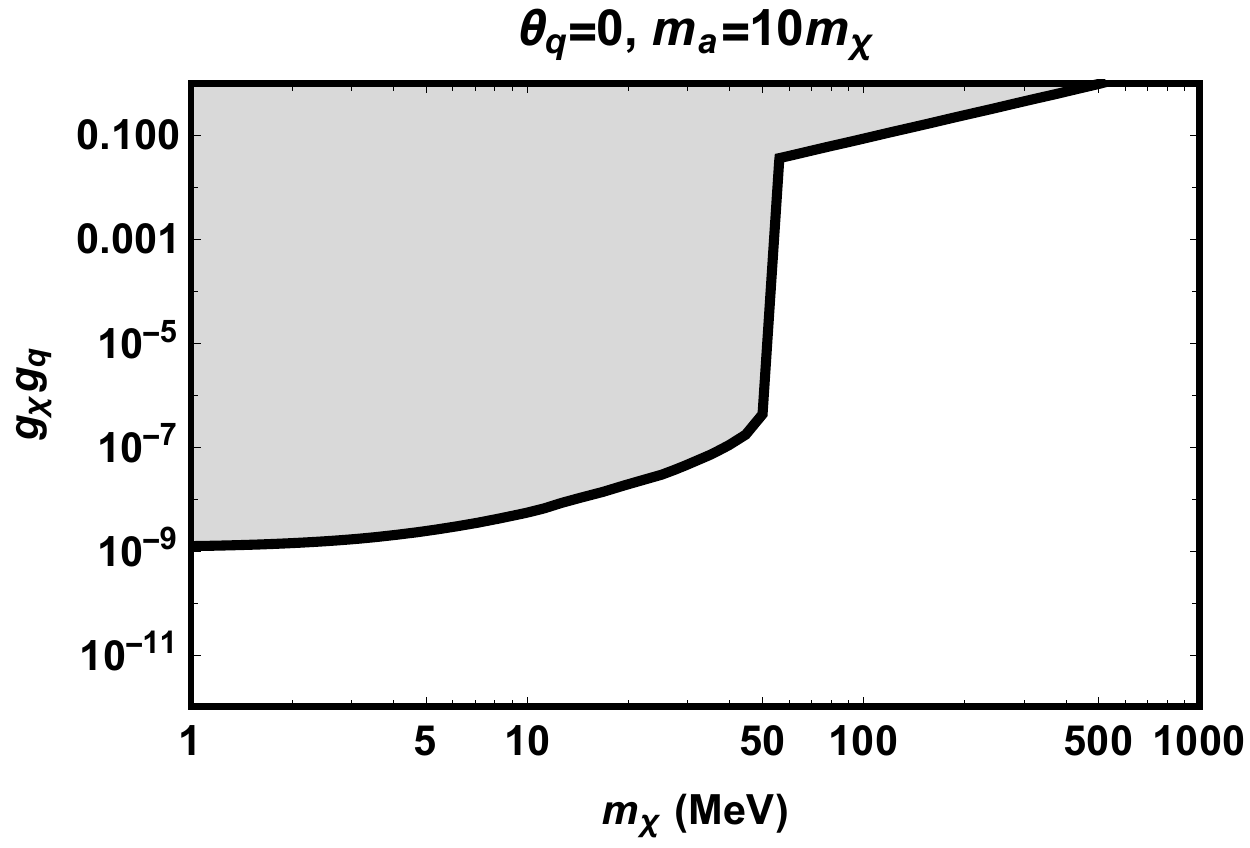}
\end{center}
\caption{The 90\% CL upper bounds on $g_\chi g_q$ as a function of the mass $m_\chi$ from COHERENT. We assume $\theta_q=0$ and $m_a=2 m_\chi$ (left), or $m_a=10 m_\chi$ (right).
}
\label{fig:thetaq=0}
\end{figure}

To obtain the bounds on the simplified neutrino model, we first set $\theta_q=0$ and $m_a=2m_\chi$ or $10m_\chi$, and scan over possible values of the product of the coefficients $g_\chi g_q$ for a given $m_\chi$. The 90\% CL upper bounds on $g_\chi g_q$ as a function of $m_\chi$ are shown in Fig.~\ref{fig:thetaq=0}. As we see from Fig.~\ref{fig:thetaq=0}, for $m_\chi \lesssim 53$ MeV, the upper bounds on $g_\chi g_q$ are very strong, and can reach as small as $10^{-9}$ for $m_\chi= 1$ MeV. For $m_\chi\gtrsim 53$ MeV, however, the tree-level process $\nu N\to \chi N$ is kinematically forbidden and the bounds become much weaker since the contribution from the loop diagrams is relatively small. Thus, there exhibits a kink around $m_\chi\simeq 53$ MeV. Compare the left panel of Fig.~\ref{fig:thetaq=0} to the right panel, we see that in general the bounds become weaker as the mediator mass increases.

\begin{figure}[htb!]
\begin{center}
\includegraphics[scale=1,width=0.48\linewidth]{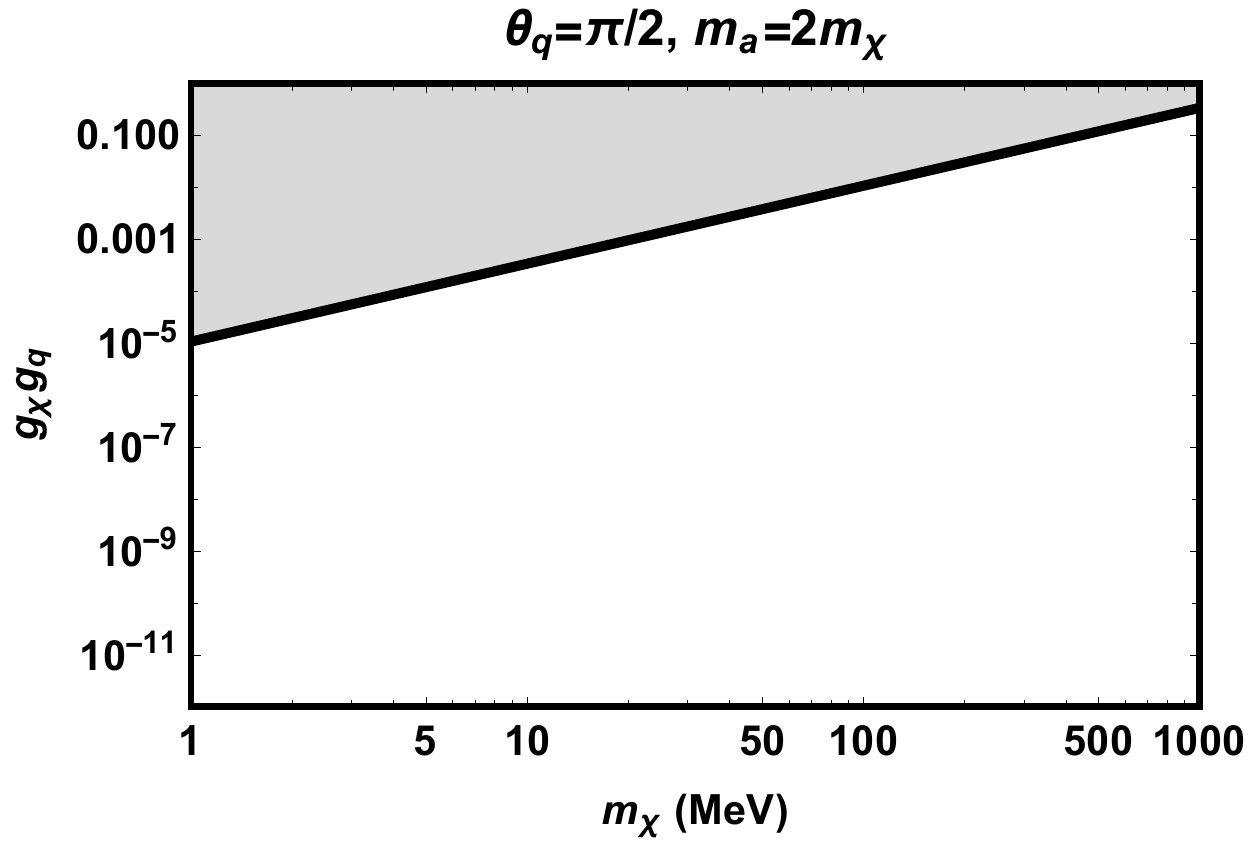}
\includegraphics[scale=1,width=0.48\linewidth]{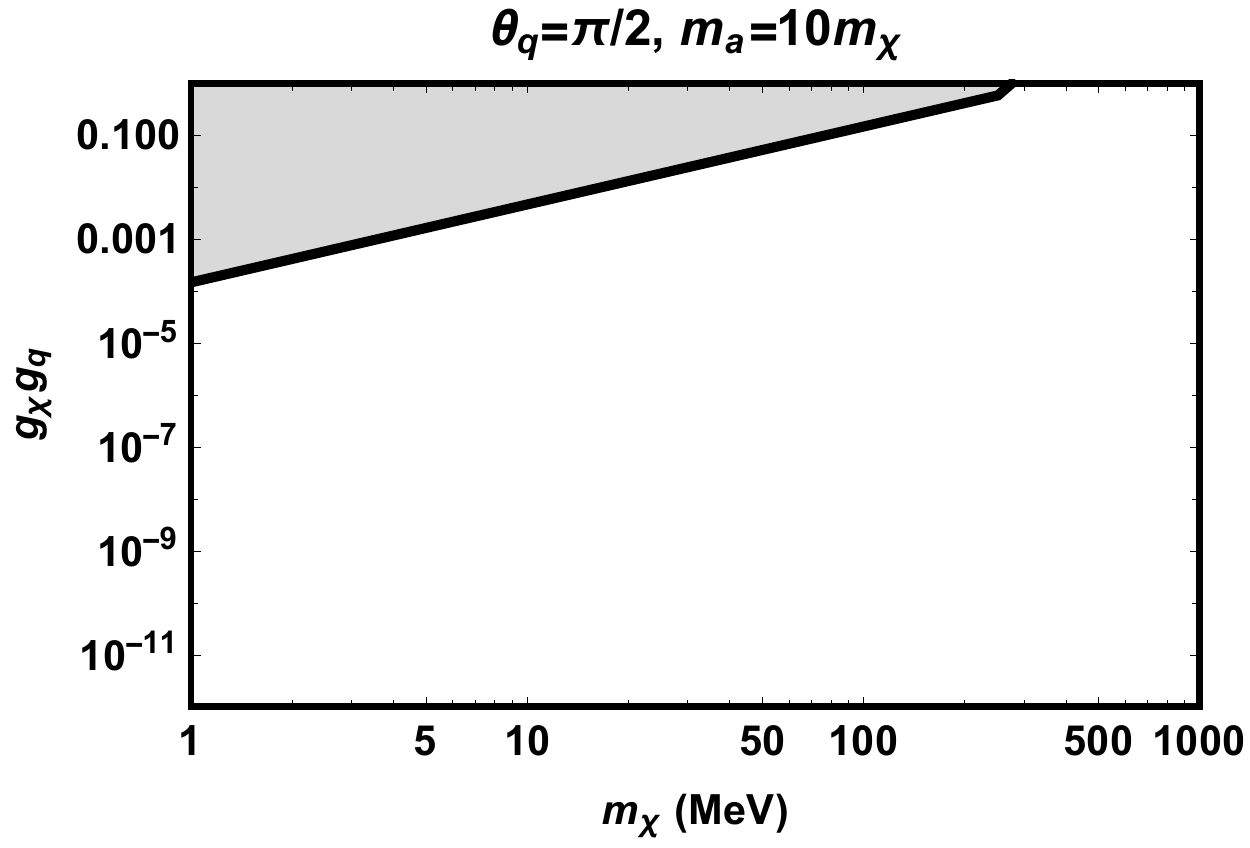}
\end{center}
\caption{Same as Fig.~\ref{fig:thetaq=0}, except for $\theta_q=\pi/2$.
}
\label{fig:thetaq=90}
\end{figure}

We also fix $\theta_q=\pi/2$ and obtain the 90\% CL upper bounds on $g_\chi g_q$ as a function of $m_\chi$. The results are shown in the left and right panel of Fig.~\ref{fig:thetaq=90} for $m_a=2m_\chi$ and $m_a=10m_\chi$, respectively. From Eq.~(\ref{eq:treelevel}), we see that for $\theta_q=\pi/2$, the SI terms from the tree-level process vanish, and the bounds are mainly determined by the loop-level contribution~\footnote{In principle, for $m_\chi\lesssim 53$ MeV, the SD terms also contribute to the CE$\nu$NS process. The contribution is nonzero for the odd-even nucleus in the CsI detector. As it is highly suppressed compared with the SI terms, we do not consider its contribution here.}. For $m_\chi\simeq 53$ MeV, the pure loop-level contribution constrains $g_\chi g_q$ to be smaller than $0.003$ ($0.06$) for $m_a=2m_\chi$ ($10m_\chi$).

\section{Other constraints}
\label{sec:Cons}

\subsection{Flavor constraint}

The simplified neutrino model receives constraints from the invisible rare decays such as $K^+\to \pi^+ + {\rm invisible}$ via flavor changing neutral currents~\cite{Dolan:2014ska}. This rare decay is recently measured by the NA62 experiment at CERN~\cite{CortinaGil:2020vlo}.
In this model the partial width for $K^+\to \pi^+ a$ is~\cite{Wise:1980ux,Leutwyler:1989xj}
\begin{eqnarray}
\Gamma(K^+\to \pi^+ a)&=& {|f(g_q)|^2+|k(g_q)|^2\over 16\pi m_{K^+}^3}\lambda^{1/2}(m_{K^+}^2,m_{\pi^+}^2,m_a^2)\;,\\
f(g_q)&=&{3m_{K^+}^2\over 32\pi^2 v_0^2 m_s}g_q\cos\theta_q f_+(0) \sum_{q=u,c,t}m_q^2 V_{qd}^\ast V_{qs}\;,\\
k(g_q)&=&{4m_{K^+}^2\over 32\pi^2 v_0^2 m_s}g_q\sin\theta_q f_+(0) \sum_{q=u,c,t}m_q^2 V_{qd}^\ast V_{qs} {\rm ln}\Big({m_W^2\over m_q^2}\Big)\;,
\end{eqnarray}
where $\lambda(x,y,z)=x^2+y^2+z^2-2xy-2yz-2xz$ and $f_+(0)\approx 0.9709$ is the vector form factor at zero momentum transfer. For $K_L\to \pi^0$ decay, one needs to replace $V_{qd}^\ast V_{qs}$ by ${\rm Re}[V_{qd}^\ast V_{qs}]$ and change the masses of kaon and pion. One can see that the contribution is from the up-type quarks in the loop and Ref.~\cite{Hurtado:2020vlj} estimated the bound as $g_q\lesssim 1.58\times 10^{-4}$.
Note that the above result is on the analogy of the derivations for the SM-like Higgs in Ref.~\cite{Leutwyler:1989xj} and the CP-odd Higgs in the two Higgs doublet model in Ref.~\cite{Wise:1980ux}. The actual calculation would suffer from a problem of UV divergence due to the fact that the simplified model here is not gauge-invariant~\cite{Dolan:2014ska,Arcadi:2017wqi}. The reliable estimate of the flavor observable relies on the UV completion realization. This problem does not affect our assumption of the couplings below and the corresponding conclusions.

The flavor observables through $K\to \pi$ transitions thus set stringent constraints on $g_q$ coupling for up-type quarks.
We assumed non-universal $g_q$ couplings for up-type and down-type quarks and neglected the up-type quark coupling in the above calculations.

\subsection{LHC constraint}

If the new fermion $\chi$ is long-lived enough, the search for events with large missing
transverse momentum with an energetic jet or the third generation SM quarks at the Large Hadron Collider (LHC) may place bounds on the coupling $g_\chi g_q$.
One can see that there is no severe bound on the very light long-lived particle of interest with spin-0 mediator from monojet search~\cite{Aaboud:2017phn,Sirunyan:2017jix} or the coupling of down-type quarks from the production in association with bottom quarks~\cite{Aaboud:2017rzf}.
The most stringent limits are from the associated production of $t\bar{t}$ with missing transverse momentum. The pseudoscalar mediator mass around 20 GeV is
excluded at 95\% confidence level, assuming the $\chi$ mass being 1 GeV and unitary top couplings~\cite{Aaboud:2017aeu}. This constraint is not severe in our discussion with absent up-type quark coupling. In addition, as stated before, the search for Higgs invisible decay sets constraint on the coupling $\lambda_{haa}\lesssim 0.01$ in a specific Higgs portal model~\cite{Khachatryan:2016vau,Khachatryan:2016whc}.

\subsection{Long-lived hypothesis}

Although we did not interpret the fermion $\chi$ as DM in the above discussion, here we briefly discuss the validation of our result on the new fermion $\chi$ if it is long-lived. In our simplified scenario, the two-body process $\chi\to \nu\gamma$ via a closed quark loop vanishes as its amplitude becomes
\begin{eqnarray}
M(\chi\to \nu\gamma)&\propto& \int {d^4 k\over (2\pi)^4} {k\cdot \epsilon(q)\over (k^2-m_q^2)[(k+q)^2-m_q^2]}\propto \epsilon\cdot q = 0\;,
\end{eqnarray}
where $\epsilon$ is the photon polarization vector, $k$ and $q$ denote the integral loop-momentum and the momentum of the external photon, respectively.
The leading decay process is thus $\chi\to \nu \gamma \gamma$ with two photons radiated from the closed quark loop. According to Ref.~\cite{Dror:2020czw}, the decay width depends on both $m_\chi$ and $g_\chi g_q\over m_a^2$
\begin{eqnarray}
\Gamma(\chi\to \nu\gamma\gamma)\propto m_\chi^7 \Big({g_\chi g_q\over m_a^2}\Big)^4 \;.
\end{eqnarray}
Therefore, if $m_\chi$ is small enough, the fermion $\chi$ can be long-lived. As seen in Fig.~\ref{fig:thetaq=0}, the constraint from COHERENT is independent of $m_\chi$ in low mass region and valid for very light fermion $\chi$. We refer a UV complete model to Ref.~\cite{Brdar:2018qqj}.

\section{Conclusions}
\label{sec:Con}

We investigate the constraint on general neutrino interactions with an exotic fermion in the coherent neutrino-nucleus scattering experiments. We consider both the tree-level and loop-level contributions to the CE$\nu$NS process in the framework of a simplified neutrino model with a new Dirac fermion $\chi$ and a spin-0 mediator $a$. The couplings between the mediator and the new fermion $\chi$ (the SM quarks) is parameterized by $g_\chi$ ($g_q$). For the CP phase in the quark sector $\theta_q=0~(\pi/2)$, the detection processes are dominated by the tree-level (loop-level) contribution.
We find that for $\theta_q=0$, the COHERENT experiment can set the upper bound of $g_\chi g_q$ as small as $10^{-9}$ for $m_\chi=1$ MeV. Also, by including the loop-level contribution, the COHERENT data is also sensitive to the mass region with $m_\chi\gtrsim53$ MeV, which is the maximal energy of the incoming neutrinos measured at COHERENT. In general, the bounds become weaker as the mediator mass increases. In addition, when $\theta_q=\pi/2$, the COHERENT spectrum can be also modified after taking the loop-level contribution into account. By fitting to the COHERENT data, we find that the loop-level contribution constrains $g_\chi g_q$ as small as $0.003$ (0.06) for $m_\chi\simeq 53$ MeV and $m_a=2m_\chi$ ($10m_\chi$).

\section*{ACKNOWLEDGMENTS}
TL is supported by the National Natural Science Foundation of China (Grant No. 11975129, 12035008) and ``the Fundamental Research Funds for the Central Universities'', Nankai University (Grant No. 63196013). JL is supported by the National Natural Science Foundation of China (Grant No. 11905299), Guangdong Basic and Applied Basic Research Foundation (Grant No. 2020A1515011479), the  Fundamental  Research  Funds  for  the  Central Universities, and the Sun Yat-Sen University Science Foundation.

\appendix

\section{Loop diagram calculation in neutrino-nucleus scattering}
\label{app:loopcalc}

For the elastic scattering process $\nu(p_1) q(k_1)\to \nu(p_2) q(k_2)$, the Passarino-Veltman function for the triangle diagram is
\begin{eqnarray}
C_0[p_1^2, (p_1-p_2)^2, p_2^2; m_\chi^2, m_a^2, m_a^2]=C_0[0,0,0; m_\chi^2, m_a^2, m_a^2]={m_\chi^2{\rm ln}\Big({m_a^2\over m_\chi^2}\Big)-m_a^2+m_\chi^2\over (m_a^2-m_\chi^2)^2}\; .
\end{eqnarray}
The Passarino-Veltman functions for the one-loop box diagrams are defined as
\begin{eqnarray}
D_{0}(p_1,m_\chi,m_a)&\equiv& D_{0}[p_1^2,p_1^2,0,0,0,p_1^2;0,m_\chi^2,m_a^2,m_a^2]={-m_a^2+m_\chi^2+m_a^2{\rm ln}\Big({m_a^2\over m_\chi^2}\Big)\over m_a^2(m_a^2-m_\chi^2)^2}\;,\\
D_{00}(p_1,m_\chi,m_a)&\equiv& D_{00}[p_1^2,p_1^2,0,0,0,p_1^2;0,m_\chi^2,0,m_a^2]-D_{00}[p_1^2,p_1^2,0,0,0,p_1^2;0,m_\chi^2,m_a^2,m_a^2]\nonumber \\
&=& {m_a^2-m_\chi^2-m_a^2{\rm ln}\Big({m_a^2\over m_\chi^2}\Big)\over 4(m_a^2-m_\chi^2)^2}\;,\\
D_{11}(p_1,m_\chi,m_a)&\equiv& D_{11}[p_1^2,p_1^2,0,0,0,p_1^2;0,m_\chi^2,0,m_a^2]-D_{11}[p_1^2,p_1^2,0,0,0,p_1^2;0,m_\chi^2,m_a^2,m_a^2]\nonumber \\
&=& m_a^2{(m_a^2-m_\chi^2)(m_a^2+5m_\chi^2)-2m_\chi^2(2m_a^2+m_\chi^2){\rm ln}\Big({m_a^2\over m_\chi^2}\Big)\over 6m_\chi^2(m_a^2-m_\chi^2)^4}\;,\\
D_{001}(p_1,m_\chi,m_a)&\equiv& D_{001}[p_1^2,p_1^2,0,0,0,p_1^2;0,m_\chi^2,0,m_a^2]-D_{001}[p_1^2,p_1^2,0,0,0,p_1^2;0,m_\chi^2,m_a^2,m_a^2]\nonumber \\
&=& m_a^2{-2m_a^2+2m_\chi^2+(m_a^2+m_\chi^2){\rm ln}\Big({m_a^2\over m_\chi^2}\Big)\over 12(m_a^2-m_\chi^2)^3}\;.
\end{eqnarray}

For the two-loop diagrams, the $\Pi_{G}(\ell^2)$ and $\Pi_{\tilde{G}}(\ell^2)$ in Eq.~(\ref{ampaaGG}) are
\begin{eqnarray}
\Pi_{G}(\ell^2)&=&\sum_{Q=c,b,t}g_Q^2\Big({m_Q\over v}\Big)^2 \int_0^1 dx \Big[ {{3\over 2}x(1-x)\over m_Q^2-\ell^2 x(1-x)} + m_Q^2 {3x(1-x)+2(-1-x+x^2)c_{2\theta_q}\over 2(m_Q^2-\ell^2 x(1-x))^2}\nonumber \\
&-&m_Q^4{1-3x+3x^2-(1-x)x c_{2\theta_q}\over (m_Q^2-\ell^2 x(1-x))^3}\Big]\;,\\
\Pi_{\tilde{G}}(\ell^2)&=&\sum_{Q=c,b,t}g_Q^2\Big({m_Q\over v}\Big)^2 \int_0^1 dx {m_Q^2 s_{2\theta_q}\over (m_Q^2-\ell^2 x(1-x))^2}\;.
\end{eqnarray}
The $F_G, F_{\tilde{G}}$ functions are
\begin{eqnarray}
F_{G}(p_1^2,m_\chi^2,m_a^2,m_Q^2)&=& \int_0^1 dx\Big[-{3\over 2}{\partial\over \partial m_a^2}X_1\Big(p_1^2,m_\chi^2,m_a^2,{m_Q^2\over x(1-x)}\Big)\nonumber \\
&+&m_Q^2{3x(1-x)+2(-1-x+x^2)c_{2\theta_q}\over 2x^2(1-x)^2}{\partial\over \partial m_a^2}X_2\Big(p_1^2,m_\chi^2,m_a^2,{m_Q^2\over x(1-x)}\Big)\nonumber \\
&+&m_Q^4{1-3x+3x^2-x(1-x)c_{2\theta_q}\over x^3(1-x)^3}{\partial\over \partial m_a^2}X_3\Big(p_1^2,m_\chi^2,m_a^2,{m_Q^2\over x(1-x)}\Big) \Big] \;,\\
F_{\tilde{G}}(p_1^2,m_\chi^2,m_a^2,m_Q^2)&=& \int_0^1 dx\Big[ m_Q^2 {s_{2\theta_q}\over x^2(1-x)^2} {\partial\over \partial m_a^2}X_2\Big(p_1^2,m_\chi^2,m_a^2,{m_Q^2\over x(1-x)}\Big) \Big] \;,
\end{eqnarray}
where
\begin{eqnarray}
X_1\Big(p_1^2,m_\chi^2,m_a^2,{m_Q^2\over x(1-x)}\Big)&=& {1\over m_a^2-{m_Q^2\over x(1-x)}}\Big[B_0(p_1^2,m_a^2,m_\chi^2)-B_0\Big(p_1^2,{m_Q^2\over x(1-x)},m_\chi^2\Big)\Big]\;,\\
X_2\Big(p_1^2,m_\chi^2,m_a^2,{m_Q^2\over x(1-x)}\Big)&=& {1\over m_a^2-{m_Q^2\over x(1-x)}}\Big[X_1\Big(p_1^2,m_\chi^2,m_a^2,{m_Q^2\over x(1-x)}\Big)-C_0\Big(p_1^2,{m_Q^2\over x(1-x)},m_\chi^2\Big)\Big]\;,\nonumber \\ \\
X_3\Big(p_1^2,m_\chi^2,m_a^2,{m_Q^2\over x(1-x)}\Big)&=& {1\over m_a^2-{m_Q^2\over x(1-x)}}\Big[X_2\Big(p_1^2,m_\chi^2,m_a^2,{m_Q^2\over x(1-x)}\Big)-D_0\Big(p_1^2,{m_Q^2\over x(1-x)},m_\chi^2\Big)\Big]\;,\nonumber \\
\end{eqnarray}
and
\begin{eqnarray}
\int {d^4\ell\over (2\pi)^4}{1\over [(\ell+p)^2-M^2](\ell^2-m^2)}&=& {i\over (4\pi)^2}B_0(p^2,m^2,M^2)\;,\\
\int {d^4\ell\over (2\pi)^4}{1\over [(\ell+p)^2-M^2](\ell^2-m^2)^2}&=& {i\over (4\pi)^2}C_0(p^2,m^2,M^2)\;,\\
\int {d^4\ell\over (2\pi)^4}{1\over [(\ell+p)^2-M^2](\ell^2-m^2)^3}&=& {i\over (4\pi)^2}D_0(p^2,m^2,M^2)\;.
\end{eqnarray}

\bibliographystyle{JHEP}
\bibliography{refs}

\end{document}